\documentclass[11pt]{article}

\usepackage{geometry}
\usepackage{currfile}
\usepackage{textcomp}

\usepackage{xcolor,soul}
\sethlcolor{yellow}

\usepackage[pdftex]{graphicx}
\DeclareGraphicsExtensions{.pdf,.jpeg,.png}
\graphicspath{{./.}}

\usepackage{fancyhdr}
\begin{document}

\title{\Large{Towards a Design Philosophy for Interoperable Blockchain Systems}\\
~~}
\author{
\large{Thomas Hardjono~~~Alexander Lipton~~~Alex ``Sandy'' Pentland}\\
~~\\
\large{MIT Connection Science}\\
\large{Massachusetts Institute of Technology}\\
\large{Cambridge, MA}\\
~~\\
}

\maketitle

\begin{abstract}
In this paper we discuss a design philosophy for interoperable blockchain systems,
using the design philosophy of the Internet architecture as the basis to identify key design principles.
Several interoperability challenges are discussed in the context of cross-domain transactions.
We illustrate how these principles are informing the interoperability architecture
of the MIT Tradecoin system.
\end{abstract}

\newpage
\clearpage

{\small 
\tableofcontents
}

~~\\

\section{Introduction}

The goal of this paper is to bring to the forefront the notion
of {\em interoperability} for blockchain systems,
using lessons learned from the three decades 
of the development of the Internet.
Our overall goal is to develop a design philosophy
for an interoperable blockchain architecture,
and to identify some design principles that
that promote interoperability.

Currently there is considerable interest (real and hype)
in blockchain systems as a promising technology for the future
infrastructure for a global value-exchange nextwork -- or what
some refer to as the ``Internet of value''.
The original blockchain idea of 
Haber and Stornetta~\cite{HaberStornetta1991,BayerHaber1993} 
is now a fundamental construct within most blockchain systems,
starting with the Bitcoin system which first adopted the idea
and deployed it in a digital currency context.

Many parallels have been made between blockchain systems and the Internet.
However, many comparisons often fail to understand the fundamental
goals of the Internet architecture as promoted and lead by DARPA.
There was a pressing need in the Cold War period of the 1960s and 1970s
to develop a new communications network architecture that did not previously exist,
one that would allow communications to survive in the face of attacks.
In Section~\ref{sec:DesignPhilosophyInternet} we review and discuss these goals.

Today, the motivations are similar but driven by different historical events.
The global financial crisis of 2008 has forced several central banks 
around the world to rethink the way they conduct monetary policy.
We argue that if blockchain technology seeks to be a fundamental
component of the future global distributed network of commerce and value,
then its architecture must also satisfy the same fundamental goals
of the Internet architecture.
In Section~\ref{sec:InteroperableBlockchainSystems}
we discuss a design philosophy for blockchain systems.
This is followed by Section~\ref{sec:tradecoin-overview}
in which we discuss the interoperability design aspects of 
the MIT Tradecoin project,
mapping much of the Internet's constructs to that
of the blockchain architecture.

\section{The Design Philosophy of the Internet}
\label{sec:DesignPhilosophyInternet}

In considering the future direction for blockchain systems generally,
it is useful to recall and understand goals of the Internet architecture
as defined in the early 1980s as a project funded by DARPA.
The definition of the Internet as view in the late 1980s
is the following:
it is ``a packet switched communications facility
in which a number of distinguishable networks
are connected together using 
packet switched communications processors
called gateways which implement 
a store and forward packet-forwarding algorithm''~\cite{Clark88,CerfKahn74}.

\subsection{Fundamental Goals}

It is important to remember that the design of the ARPANET and the Internet favored
military values (e.g. survivability, flexibility, and high performance)
over commercial goals (e.g. low cost, simplicity, or consumer appeal)~\cite{abbate99},
which in turn has  affected how the Internet has evolved and has been used.
This emphasis was understandable given the Cold War backdrop to the 
packet-switching discourse throughout the 1960s and 1970s. 

The DARPA view at the time was that there are seven (7) goals of the Internet architecture,
with the first three being fundamental to the design,
and the remaining four being second level goals.
The following are the fundamental goals of the Internet in the order of importance~\cite{Clark88,CerfKahn74}:
\begin{enumerate}

\item	{\em Survivability}: Internet communications must continue despite loss of networks or gateways.

This is the most important goal of the Internet,
especially if it was to be the blueprint for military packet switched communications facilities.
This meant that if two entities are
communicating over the Internet, and some failure
causes the Internet to be temporarily disrupted and
reconfigured to reconstitute the service, then the entities
communicating should be able to continue without
having to reestablish or reset the high level state of their conversation.
Therefore to achieve this goal, the state information which
describes the on-going conversation must be protected.
But more importantly, in practice this explicitly meant that
it is acceptable to lose the state
information associated with an entity if, at the same
time, the entity itself is lost.
This notion of state of conversation
is related to the end-to-end principle discussed below.

\item	{\em Variety of service types}: The Internet must support multiple types of communications service.

What was meant by ``multiple types'' is that at the transport level
the Internet architecture should support
different types of services
distinguished by differing requirements for
speed, latency and reliability.
Indeed it was this goal that resulted
in the separation into two layers
of the TCP layer and IP layer,
and the use of bytes (not packets) at the TCP layer
for flow control and acknowledgement.

\item	{\em Variety of networks}: The Internet must accommodate a variety of networks.

The Internet architecture must be able
to incorporate and utilize a
wide variety of network technologies, including military
and commercial facilities.

\end{enumerate}

The remaining four goals of the Internet architecture are:
(4) distributed management of resources,
(5) cost effectiveness,
(6) ease of attaching hosts,
and
(7) accountability in resource usage.
Over the ensuing three decades these second level goals
have been addressed in different ways.
For example, accountability in resource usage evolved
from the use of rudimentary management information bases (MIB)~\cite{rfc1066,rfc1156}
into the current sophisticated traffic management protocols and tools.
Cost effectiveness was always an important aspect of the business model for consumer ISPs
and corporate networks.

\subsection{The End-to-End Principle}

One of the critical debates throughout the development of the Internet architecture
in the 1980s was in regards to the placement of functions
that dealt with reliability of message delivery
(e.g. duplicate message detection, message sequencing, guaranteed message delivery, encryption).
In essence the argument revolved around 
the amount of effort put into reliability measures within the data
communication system, and was seen as 
an engineering trade-off based on performance.
That is, how much low-level function (for reliability) needed to be implemented
by the networks versus implementation by the applications at the endpoints.

The line of reasoning against low-level function implementation in the network became known
as the {\em end-to-end argument} or principle.
The basic argument is as follows: a lower level subsystem that supports a distributed
application may be wasting its effort in providing a function that must
be implemented at the application level anyway~\cite{SaltzerReed84}.
Thus, for example, for duplicate message suppression the task
must be accomplished by the application itself seeing that the application
is most knowledgeable as to how to detect its own duplicate messages.

Another case in point relates to data encryption.
If encryption/decryption was to be performed by the network, 
then the network and its data tranmission systems
must be trusted to securely manage the
required encryption keys.
Also, when data enters the network (to be encrypted there) the data
will be in plaintext and therefore susceptible to theft and attacks.
Finally, the recipient application of the encrypted messaged will still need
to verify the source-authenticity of the message.
The application will still need to perform key management.
As such, the best place to perform data encryption/decryption
is the application endpoints -- there is no need for the
communication subsystem to provide for automatic encryption of all traffic.
That is, encryption is an end-to-end function.

The end-to-end principle was a fundamental design principle
of the security architecture of the Internet.
Among others, it influenced the direction of the subsequent
security features of the Internet,
including the the development
of the IP-security sublayer~\cite{rfc2401}
and its attendant key management function~\cite{rfc2408,rfc2409}.
Today the entire Virtual Private Network (VPN) subsegment of the networking industry
started based on this end-to-end principle.
(The global VPN market alone is forecasted to reach 70 billion dollars in the next few years).
The current day-to-day usage of the Secure Sockets Layer (TLS)~\cite{rfc2246}
to protect HTTP web-traffic (i.e. browsers) is also built on the premise
that client-server data encryption is an end-to-end function
performed by the browser (client) and by the HTTP server.

\subsection{The Autonomous Systems Paradigm}

A key concept in the development of the Internet is that
of {\em autonomous systems} (or {\em routing domains})
as a connectivity unit that provide scale-up capabilities.
The notion of autonomous systems provides
a way to {\em hierarchically aggregate routing information},
such the distribution of routing information itself becomes
a manageable task.
This division into domains provides independence from each domain owner/operator
to employ the routing mechanisms of its own choice.
IP packet routing inside an autonomous system
is therefore referred to as {\em intra-domain} routing,
while routing between (across) autonomous systems
is referred to as {\em inter-domain} routing.
The common goal of many providers of routing services
(consumer ISPs, backbone ISPs and participating corporations)
is that of supporting different types of services
(in the sense of speed, latency and reliability).

In the case of routing
the aim is to share best-route information among routers using
an intra-domain routing protocol (e.g. distance vector such as RIP~\cite{rfc2453},
or link-state such as OSPF~\cite{rfc2328}).
The routing protocol of choice must address numerous issues,
including possible loops and imbalances in traffic distribution.
Today routers are typically owned and operated by the legal owner
of the autonomous system (e.g. ISP or corporation).
These owners enter into {\em peering} agreements
with each other in order to achieve end-to-end reachability
of destinations across multiple hops of domains.
The primary revenue model in the ISP industry revolves
around different tiers of services appropriate to
different groups of customers.

\begin{figure}[!t]
\centering
\includegraphics[width=0.9\textwidth, trim={0.0cm 0.0cm 0.0cm 0.0cm}, clip]{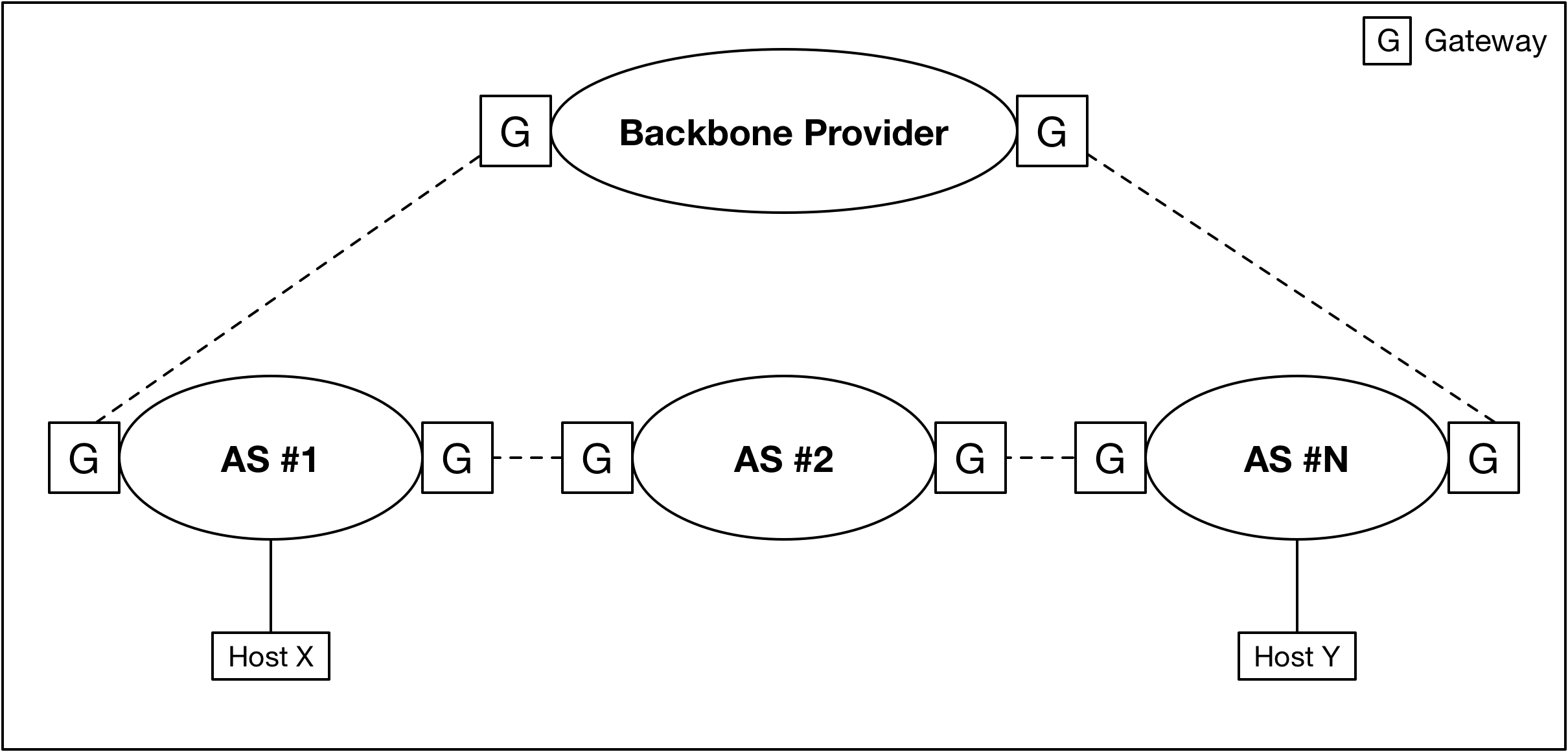}
	%
	%
\caption{Autonomous Systems as a set of networks and gateways (after~\cite{CerfKahn74})}
\label{fig:AutonomousSystems}
\end{figure}

There are several important points regarding autonomous systems 
and the positive impact this paradigm has had on the 
development of the Internet for the past several decades:
\begin{itemize}

\item	{\em Autonomous systems paradigm leads to scale and reach}:~\\
The autonomous system paradigm, the connectionless routing model and 
the distributed network topology
of the Internet
allows each unit (the AS) to solve performance issues locally.
This in turn promotes service scale in the sense of throughput (end-to-end)
and reach (the large numbers of connected endpoints).
As such, it is important to see the global Internet today a connected
set of ``islands'' of autonomous system,
stitched together through peering agreements.

\item	{\em Domain-level control with distributed topology}:~~\\
Each autonomous system typically possesses multiple
routers operating the same intra-domain routing protocol.
The availability of multiple routers implies availability of
multiple routing paths through the domain.
Despite this distributed network topology,
these routers are centrally controlled
(e.g. by the network administrator of the domain).
The autonomous system as a control-unit provides
manageability, visibility and peering capabilities
centrally administered by the owner of the domain.

\item	{\em Each entity is uniquely identifiable in its domain}:~\\
All routers (and other devices, such as bridges and switches) in an autonomous system
are uniquely identifiable and visible to the network operator.
This is a precondition of routing.
The identifiability and visibility of devices in a domain is usually limited to that domain.
Entities outside the domain may not even be aware of the existence
individual routers in the domain.

\item	{\em Autonomous system reachability}:
Autonomous systems interact with each other through
special kinds of routers (called ``Gateways'') that are designed and configured
for cross domain packet routing.
These operate specific kinds of protocols (such as
an exterior Border Gateway Protocol~\cite{rfc1105,rfc4271}),
which provides transfer packets across domains.
For various reasons (including privacy and security)
these exterior-facing gateway protocols advertise only {\em reachability}
status information regarding routers and hosts in the domain.
They do not make the routing conditions in the domain and other domain throughput
information visible to external entities (i.e. other autonomous systems).
This limited visibility prevents external domains
from performing traffic shaping that may adversely impact
a given autonomous system (e.g. driving too much traffic to a given domain).

\item	{\em Autonomous systems are owned and operated by legal entities}:~\\
All routing autonomous systems (routing domains) today
are owned, operated and controlled by known entities.
Internet Service Providers (ISPs) provide their
{\em Autonomous System Numbers} (ASNs) and routing prefixes
to {\em Internet Routing Registries} (IRRs).
IRRs can be used by ISPs to develop routing plans.
An example of an IRR is 
the American Registry for Internet Numbers (ARIN)~\cite{ARIN-website},
which is one of several IRRs around the world.

\end{itemize}

It is worthwhile pointing out that the pioneers of the Internet understood the importance
of identity, and certainly did not forget it -- as is claimed by some today.
Indeed several groups in the IETF discussed the various issues
around identity, digital certificates and PKI (e.g. SPKI~\cite{rfc2693}, X.509~\cite{rfc2459}, PGP~\cite{rfc1991})
on-going for several years.
It was clear even in the mid-1990s that user (human) identity
was a function to be provided at layers above the IP routing layer
and was a construct extraneous to routing packets.

\section{Interoperable Blockchains: Towards a Design Philosophy}
\label{sec:InteroperableBlockchainSystems}

In this section we use the fundamental goals of the Internet architecture
in the context of interoperable blockchain systems.
In the 1980s there was a clear realization 
that it was necessary to incorporate the then existing network architectures if
Internet was to be useful in a practical sense. 
These networks represent administrative boundaries of
control~\cite{Clark88}.
Today we are seeing a similar situation,
in which multiple blockchain designs are being proposed,
each having different degrees of technological maturity.

Different organizations and consortiums
(e.g. R3/Corda~\cite{R3-website}, EEA~\cite{EEA-website})
are developing different blockchain technologies.
Additionally, several dozen digital currencies
are operating today,
and several digital currency exchanges have emerged.
A critical aspect of these proposals is their need to address 
the fundamental question of {\em privacy} of transacting parties.
Some designs (e.g. Hyperledger Fabric~\cite{AndroulakiBarger2018})
recognize the primacy of privacy and security,
and address these question through a permissioned design.
Others are pursuing a permissionless model,
using advances cryptographic techniques (e.g. homomorphic encryption;
multi-party computation) as a ``layer'' atop the public permissionless blockchain.

We believe the issue of survivability to be as important
as that of privacy and security.
As such, we believe that interoperability across blockchain systems
will be a core requirement -- both at the mechanical level and the value level --
if blockchain systems and technologies are to become
fundamental infrastructure components of future global commerce.

In this section we identify and discuss
some of the challenges to blockchain interoperability,
using the Internet architecture as a guide
and using the fundamental goals as the basis for
developing a design philosophy for interoperable blockchains.
We offer the following definition of an
``interoperable blockchain architecture'' 
using the NIST definition of ``blockchain'' (p.50 of~\cite{NIST-80202}):
\begin{quote}
{\em
An interoperable blockchain architecture
is a composition of distinguishable blockchain systems,
each representing a distributed data ledger,
where transaction execution may span multiple blockchain systems,
and where data recorded
in one blockchain is reachable and verifiable
by another possibly foreign transaction in a semantically compatible manner.
}
\end{quote}

\begin{figure}[!t]
\centering
\includegraphics[width=0.9\textwidth, trim={0.0cm 0.0cm 0.0cm 0.0cm}, clip]{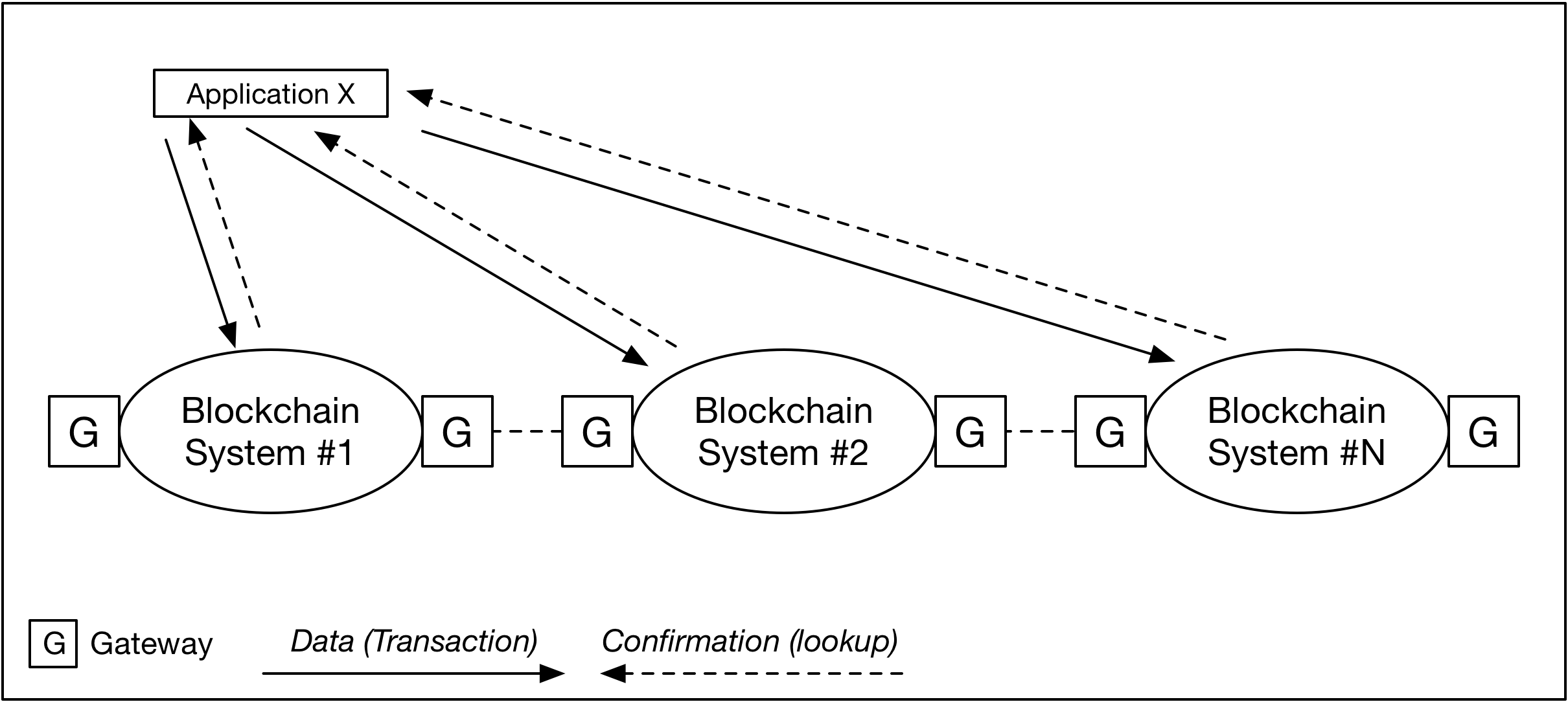}
	%
	%
\caption{Example of the reliability of a simple transaction}
\label{fig:bcsurvivability}
\end{figure}

\subsection{Survivability}
\label{subsec:Survivability}

The idea that a blockchain system
can withstand a concerted attack 
simply because it consists of physically distributed nodes
is an untested and unproven proposition.
The possible types of attacks to a blockchain system has been discussed elsewhere,
and consists of a broad spectrum.
These range from classic network-level attacks 
(e.g. network partitions; denial of services; etc.) 
to more sophisticated attacks targeting the particular constructs 
(eg. consensus implementation~\cite{EyalSirer2014,GarayKiayias2015,GervaisKarame2014}),
to targeting specific implementations of mining nodes (e.g. code vulnerabilities; viruses).

For blockchain systems we re-interpret the term ``survivability'' in the sense of the completion of
an application-level transaction.
Completion here means from its transmission
by an end-user application 
(or by a smart contract on an origin blockchain)
to its confirmation on a single blockchain system or multiple systems.
The application-level transaction may be composed of multiple
ledger-level transactions (sub-transaction) and 
which may be intended for multiple distinct blockchain systems
(e.g. sub-transaction for asset transfer, 
simultaneously with sub-transaction for payments
and sub-transaction for taxes).
Thus the notion of packets routing through multiple domains
being opaque to the communications application (e.g. email application, browser)
is re-cast to the notion of 
{\em sub-transactions confirmed on a spread of blockchain systems
being opaque to the user application}.

In the blockchain case reliability and ``best effort delivery'' 
becomes the challenge of ensuring that
an application-level transaction is completed
within reasonable time,
possibly independent of the actual blockchains where
different ledger-level sub-transactions are completed and confirmed.
Thus, the notion of `` connectionless'' here means that
in a two party application-level engagement
both parties should not care which specific blockchain systems
confirmed their sub-transaction,
so long as all confirmed sub-transactions add up
to a confirmed application-level transaction.

To illustrate the challenges of survivability,
we start with the simplest case in which
an application sends the ``data'' (signed hash value) to a blockchain for the purpose
of recording it on the ledger of the blockchain.
We ignore for the moment the dichotomy of permissionless and permissioned
blockchains and ignore the specific logic syntax of the blockchain.
Here the application does not care {\em which} blockchain records the data
so long as once it is recorded the application
(and other entities) can later find the transaction/block
and verify the data has been recorded.
Some form of confirmation must be made available
by the blockchain to the transmitting application
(e.g. it shows up on a confirmed block).

Figure~\ref{fig:bcsurvivability} illustrates the scenario.
The application transmits data-bytes (hash) to a blockchain system {No.~1},
and waits for confirmation (of the successful recording to the ledger)
to become available.
After waiting for some predetermined time unsuccessfully (i.e. timeout),
the application 
transmits the same data-bytes to a different blockchain system {No.~2}.
The application continues this process until
it is able to obtain the desired confirmation.
Thus for the application ``survivability'' means that 
the simple transaction has been successfully confirmed on some blockchain
(i.e. on all nodes of that ledger),
even if soon after the blockchain ceases being able
to process future transactions due to attacks.
The side effect maybe that the same
transaction is confirmed independently on multiple blockchain systems,
but the application does not care about this possible side effect
so long as ``the transaction got through''.

Although Figure~\ref{fig:bcsurvivability} may appear overly simplistic,
it brings forth a number of questions which echo those posed in the early days of the Internet
architecture development:
\begin{itemize}

\item	{\em Application degree of awareness}: 
To what degree must an application be aware 
of the internal constructs of a blockchain system 
in order to interact with it and make use of the blockchain.

As a point of comparison, an email client application today 
is not aware of constructs of packets, MPDUs, routing and so on.
It interacts with mail-server according to a high-level protocol
(e.g. POP3, IMAP, SMTP) and a well-defined API.

\item	{\em Placement of functions dealing with reliability}: 
What is the correct notion of ``reliability'' in the context of interoperable blockchain systems
and where should the 
function of reliability be placed.
That is, should the function of re-transmitting the same data-bytes (transaction)
be part of the application, part of the blockchain system
or part of a yet to be defined ``middle layer''.

In the case of the Internet architecture,
reliability of transmission is provided by the TCP protocol,
which has a number of flow control features
that ``hides'' reliability issues from the higher level applications.

\item	{\em Semantic type of blockchain}:
What mechanism is needed to communicate to an external application the semantic type
of the ledger-level transaction supported by a given blockchain system.
For example, a blockchain system for payments is 
different from one for recording assets,
and furthermore
different payments blockchains around the world may be implemented differently.
Merely publishing an application-level API
does not guarantee interoperability at the blockchain level.

\item	{\em Distinguishability of blockchain systems}:
For an interoperable blockchain architecture,
how does an application distinguish among blockchain systems
(even if they have compatible semantics)
and at what level should an application be aware.

Assuming the existence of multiple blockchain systems
that can serve the need of the application in Figure~\ref{fig:bcsurvivability},
how does the application distinguish between these blockchain systems.
Furthermore, should the application even need know
the actual blockchain system that successfully completed the request.

\item	{\em Objective benchmarks for speed and throughput}:
How do external entities obtain information about 
the current performance of a blockchain system
and what measure can be used to compare across systems.

\end{itemize}

One of the key considerations in the design of the Internet architecture
is the real possibility in the case of emergencies
for private networks to be temporarily placed
under government control for the purposes of government/military communications.
The interoperability of networks was therefore crucial in answering this need.
Similarly, today the question applies to blockchain systems.
In the case of emergencies
could independent and/or private blockchain systems
be temporarily placed under government control
such that relevant transactions (e.g. central bank transactions) 
can continue to flow.
The interoperability of blockchain systems is crucial in answering this future need.

\subsection{Variety of service types}

The second goal of the Internet architecture was the support
for different types of services,
distinguished by different speeds, latency and reliability.
The bi-directional reliable data delivery model
was suitable for a variety of ``applications'' on the Internet
but each application required different speeds
and bandwidth consumptions
(e.g. remote login; file transfer, etc).
This understanding led to the realization
early in the design of the Internet
that more than one transport service would be needed
and that the architecture must support
simultaneously transports wishing
to tailor reliability, delay or bandwidth usage.
This resulted in the separation of TCP
(that provided reliable sequenced data stream)
from the IP protocol that provided ``best effort''
delivery using the common building block
of the {\em datagram}.
The User Datagram Protocol (UDP)~\cite{RFC768}
was created to address the need for certain applications
that wished to trade reliability for 
direct access to the datagram construct.

For interoperable blockchain systems,
we re-interpret the goal of supporting a variety of services
to supporting the following variety of transaction aspects:
(i) speed and achieved-majority of confirmation of a given system;
(ii) the directionality of transactions;
(iii) the strength of consensus:

\begin{itemize}

\item	{\em Speed and achieved majority}:
The speed (or ``throughput'') of a blockchain system
refers to the confirmation speed,
based on the population size of the participating nodes
and other factors.

\item	{\em Directionality of ledger-transactions}: The directionality 
refers to whether the transmitting application is acting alone
or in a request-response mode of engagement with a second party
(or a group).

	\begin{itemize}
	\item	{\em Uni-directional transactions}: A transaction is uni-directional
	if the transmitting application does not intend it for any specific entity and
	no response from a peer application is expected.
	An example wold be a simple asset registry blockchain
	which records a hash of the digital scan
	of certificates (e.g. shares; land deeds; etc).

	\item	{\em Bi-directional transactions}: A transaction is bi-directional
	if the transmitting application intends that transaction for a peer application
	and expects a response-transaction from that peer.
	\end{itemize}

\item	{\em Strength of consensus}: An important consideration
for users and applications seeking to refer to (and therefore rely on)
data recorded on a ledger within a blockchain system
is the size of the population of nodes 
(i.e. entities contributing to the consensus) at any given moment
and whether this information is obtainable.
Obtaining this information maybe challenging
in systems where nodes are either anonymous,
or perhaps unobtainable by external entities in the case
of permissioned systems.

In the case of smart contracts,
there is also the question regarding the {\em provenance}
and source-authenticy
of the (externally-sourced) data being used by 
a smart contract to compute an outcome.

\end{itemize}

\begin{figure}[!t]
\centering
\includegraphics[width=1.0\textwidth, trim={0.0cm 0.0cm 0.0cm 0.0cm}, clip]{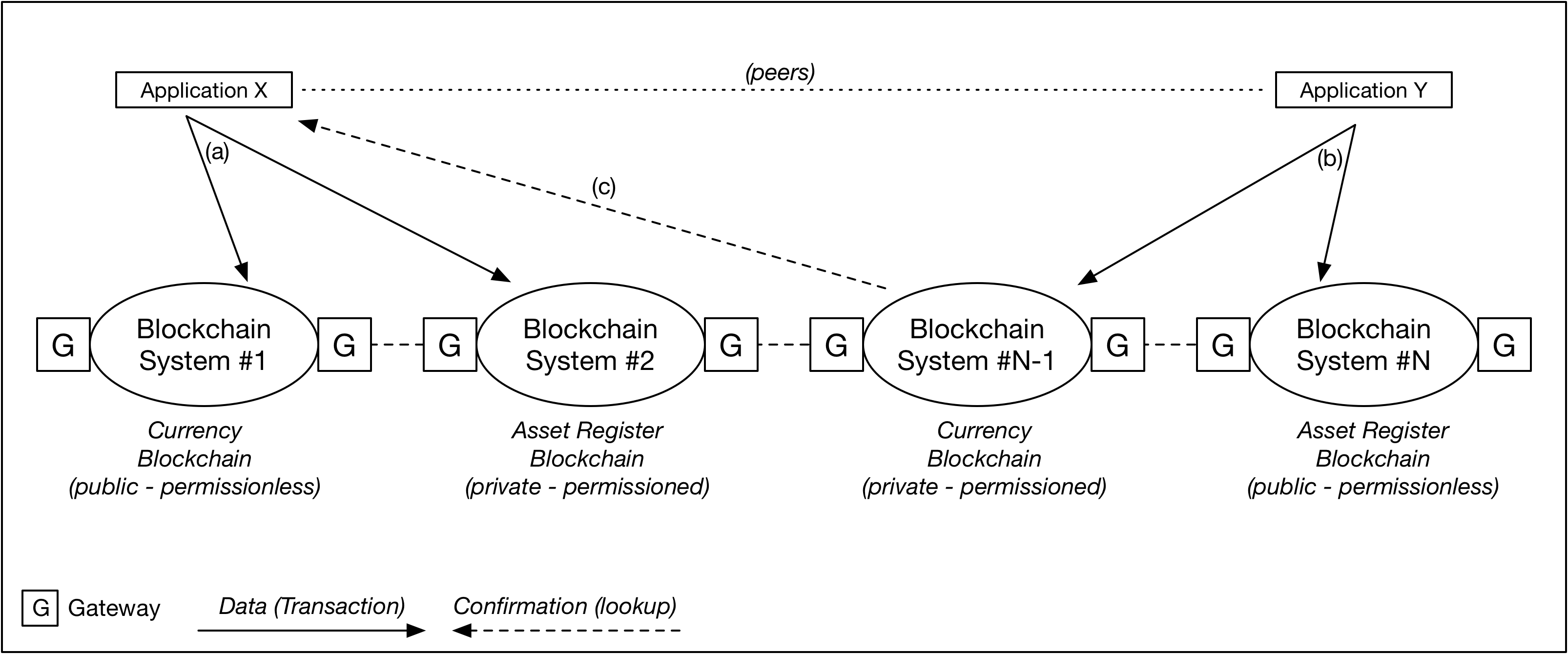}
	%
	%
\caption{Example of peer applications using different variety of blockchains systems}
\label{fig:twoapplications}
\end{figure}

\subsection{Variety of blockchain systems}
\label{subsec:varietybc}

The third fundamental goal of the Internet architecture
was to support a variety of networks,
which included networks employing different transmission technologies (e.g. {X.25}, SNA, etc.),
local networks and long-haul networks,
and networks operated/owned by different legal entities.
The {\em minimum assumption} of the Internet architecture
-- which is core to the success of the Internet as an interoperable system of networks --
was that each network must be able to transport a datagram or packet
as the lowest unit common denominator.
Furthermore, this was to be performed ``best effort''
-- namely with reasonable reliability,
but not perfect reliability.

From a blockchain interoperability perspective,
one possible re-interpretation of this original
problem is as follows: 
how can multiple types of blockchain systems
support the completion of a bi-directional transaction
between two applications,
involving computational resources across blockchain systems
where some maybe operated (or owned) by different entities.
Figure~\ref{fig:twoapplications} illustrates
a simplistic example.

In Figure~\ref{fig:twoapplications} applications X and Y
are each employing different blockchain systems
relating to currency/payments and asset ownership.
Each blockchain system implements
a different semantic logic and 
each operates under a different permissioning regime.
When application X seeks to interact with foreign application Y,
each may not have sufficient privileges to read from
the permissioned blockchain where their previous transactions have been confirmed.
Thus, when application X wishes to transfer (to Y) asset ``ownership'' 
(e.g. land deed) currently in blockchain system {No.~1} (permissioned),
application Y has no way to validate the ownership of the asset.
This is because the foreign application Y does not have authorization to read 
from the ledger in blockchain system {No.~1}.
This problem is further compounded in the case of smart contracts
that incorporate parts of the business logic of the applications.

As such, this dilemma raises several questions,
including one pertaining to the minimum assumption
for interoperable blockchain systems:
\begin{itemize}
\item	{\em Minimal assumption}: What is the minimal assumption
for interoperable blockchain systems
with regards to the notion of transaction units.
In other words, what is the ``datagram'' equivalent of transactions --
namely the transaction unit that is semantically understandable
(processable) by multiple different blockchain systems.

\item	{\em Degrees of permissionability}:
Currently the permissionless/permissioned distinction refers
to the degree to which users can participate in the system~\cite{NIST-80202}.
Interoperability across permissioned blockchains poses additional
questions with regards to how data recorded on the ledger
can be referenced (referred to or ``pointed to'') by
transactions in a foreign domain (i.e. another blockchain system).

\item	{\em Degrees of anonymity}:
There are at least two (2) degrees of anonymity that is relevant
to blockchain systems.
The first pertains to the anonymity (i.e. identity-anonymity~\cite{Chaum81,Brands93a})
of the users and the second to that of the nodes
participating in processing transactions
(e.g. nodes participating in a given consensus instance).

Combinations are possible,
such as where a permissioned system
may require all consensus nodes to be strongly authenticated and identified,
but allows for end-users to remain permissionless (and even unidentified/unauthenticated).

\end{itemize}

\subsection{Reachability}

Peering agreements between autonomous systems allows
a patchwork of islands of autonomous systems to collectively
provide routability of packets from its ingress point to its destination.
Gateway routers (entities labelled $G$ in Figure~\ref{fig:AutonomousSystems}) 
do not export full routing information or paths to entities
external to the  autonomous system.
Instead, gateway devices only provide {\em reachability} information
regarding local hosts or other autonomous systems accessible through its domain.
This model is partly driven by business survivability,
where competing ISPs (under a peering agreement) may 
purposely ``push'' (offload) traffic towards competitor ISPs,
using traffic shaping tools and algorithms.

A model akin to this gateway approach may be suitable for
the interoperability of blockchain systems.
Using the example in Figure~\ref{fig:twoapplications},
the gateways in blockchain system {No.~2} (permissioned)
can act as a proxy for the permissioned blockchain.
When a non-permissioned application Y
seeks to obtain information regarding confirmed transactions
in blockchain {No.~2},
it must present a {\em delegated} authorization from
application X who has access privileges to blockchain {No.~2}.
We discuss delegation in Section~\ref{subsec:RoleGateways}.

\subsection{Interconnecting Values}

The architecture of the Internet was a messaging delivery architecture
that separated the mechanical transmission of packets
from the {\em value} of the information contained in the packets.
The notion of priority packets was supported,
but it was primarily intended for control data
versus content payload data.
As such, the notion of value was external to the Internet.
Today this model remains also true for the majority blockchain systems.
For example, the Bitcoin system~\cite{Bitcoin} does not bind the BTC currency denomination
to any real-word asset,
and a such the notion of value is an external one (``in the eye of the beholder'').

For specific families of applications, such as currency and financial applications,
the ability to transfer value from one system to another 
is paramount and indeed the sole purpose of those applications and systems.
Thus for a semantically homogenous or near-homogenous network of blockchain systems
(e.g. payments) the challenges of value-transferral becomes
more manageable.

A promising direction in this respect is 
the {\em Inter-Ledger Protocol} (ILP) proposal~\cite{ILP2016}
which puts forward a packet format and per-hop transferal protocol
to transmit value (payments) from a sender to a receiver
over a network of currecy blockchain systems.
The end-to-end behavior of ILP
is reminiscent of the Resource Reservation Protocol (RSVP)~\cite{rfc2205}
in which a bi-directional path is ``reserved'' from the origin
to destination, and where the path needs to exist
for a short duration of time.
To do this the RSVP protocol reserves in/out interfaces (and other computational resources)
at each router per-hop from the origin to destination.

In the ILP model the sender and receiver of the payment 
are assumed to be on distinct blockchain systems.
The ILP architecture employs a value {\em connector} at the application level
between two blockchain segments.
Thus at each hop through the path
from sender to receiver there may be a per-hop connector deployed.
The function of the connector entity is to perform value-conversion
from one currency to another.
The connector behaves very similarly to a currency-exchange,
and therefore a connector entity must
have sufficient reserves of ``foreign'' currencies
(for each currency it supports)
in order to participate in the path being formed
from the sender to the receiver.
The connector model also mimics the behavior of routers and gateways
in dealing with overload to their interfaces.
In the case of ILP a connector becomes overloaded when
most or all of its pair-wise denominations
have been used or ``reserved'' in one or more
payment paths through that connector.
In this case an overloaded connector can simply reject
new requests until some of its open paths have been
closed and its resources (denominations) have become freed-up (settled).

It is important to note that the connector in ILP
represents the value-points
which are external to the blockchain systems involved.
That is, the notion of value remains
separated from the transaction mechanics of the underlying
blockchain systems.

\subsection{Moveable Smart Contracts for Survivability}

For many observers and user of blockchain technology, 
the potentially revolutionizing aspect of
blockchain technology is not so much the immutability of the ledger but instead
the notion of {\em smart contracts},
which are invokable executable-code present on P2P nodes
of the blockchain system~\cite{NortonRoseFulbright2016}.
The concept of smart contracts resembles `stored procedures'' (in classic database systems),
but in the case of smart contracts they would reside on many or all nodes
of the blockchain.
However, in the case of unreachable blockchain systems
(e.g. one under attack and therefore has a degraded throughput),
the resident smart contracts may not be invokable
or may not be able to complete/terminate due to severe resource shortages.

The following are some challenges related to survivability
of the smart contract feature when a blockchain is under attack:
\begin{itemize}
\item	Is there a {\em minimal common syntax} for smart contracts that allow
a smart contract to be copied (``moved'') from nodes in one blockchain to nodes
at a different blockchain system, where execution in the new blockchain yields
identical or semantically-equivalent output
(both technically and legally).

\item	Should the physical location of execution (i.e. which blockchain system)
of smart contracts be opaque to applications.

\item	How can an application waiting too long (timing-out) for a smart contract
in one blockchain trigger or initiate the moving (copying) of the smart contract
to a new blockchain system.

\end{itemize}

\subsection{Cryptographic Survivability}

An important consideration with regards to blockchain-based proposals
is the complexity of the ``cryptographic schemes'' underlying
some proposed blockchain systems.
Many cryptographic schemes are composed
of multiple cryptographic building blocks.
In each scheme some components maybe well understood, field deployed
and even standardized.
However, other components may represent a new idea that have never been 
field tested or withstand theoretical or practical attacks.
The point here is that the development of cryptographic technology
is a Darwinian evolutionary process,
in which a successful attack one day
becomes lessons learned for improvements for the next design.

A good case in point is the evolutionary process
undergone by symmetric ciphers (block ciphers)
for the past three decades (e.g. DES, DES3, AES, etc.).
Symmetric ciphers are a crucial component not only
of national defense infrastructures,
but also of the global commercial banking industry.
Similarly, weaknesses found in some asymmetric ciphers (e.g. RSA)
often results in recommended key lengths being extended
(e.g. NIST SP 800-175B).

Today some commercial blockchain systems are proposing to use 
complex schemes (e.g. Zero-Knowledge Proofs (ZKP), homomorphic encryption, etc.),
many of which are still at research stage at best.
It remains to be seen how attacks on these schemes
impact the blockchain systems which employ them.

\section{Interoperability Design in Tradecoin}
\label{sec:tradecoin-overview}

The MIT Tradecoin project~\cite{TradecoinRSOS2018} 
has a number of objectives,
one core goal being the development of a ``blueprint'' model
for interoperable blockchain systems which
can be applied to multiple use-cases.
Some uses cases which have been identified are: 
(i) a reserve digital currency shared by a number of geopolitically diverse small countries,
as a means to provide local financial stability~\cite{TradecoinSciAm2018};
(ii) a digital currency operating for a narrow-bank that can
provide relative stability during financially volatile periods~\cite{TradecoinCapco2018};
(iii) farmers with crop assets that wish to combine 
their assets to achieve better market presence;
(iv) logistics chains with many cooperating companies holding 
assets that will be combined into a final product or service.
As a blue-print, the Tradecoin interoperability model should be 
independent of any specific blockchain implementation.

In this section we discuss various aspects of the Tradecoin interoperability model,
following the notion of autonomous systems
as developed in the Internet architecture.
Attention is given specifically
to cross-domain transactions and the role of ``gateways'' (special nodes or computers)
that support interoperability across blockchain autonomous systems.

Although this section focuses on the technical aspects of gateways,
it is generally understood that
interoperability needs to occur both
at the technical (mechanical) level
and at the ``value'' level:
\begin{itemize}

\item	{\em Mechanical level interoperability}: This layer encompasses the computer
and network systems 
(hardware and software) that implement the technical blockchain constructs
as well as the communications constructs.
This layer contains protocols, cryptography, encryption, signing,
identities (identifiers), operational governance rules,
consensus algorithms, transactions, probes and so on.

\item	{\em Value level interoperability}: 
This layer is external to the blockchain system
and encompasses constructs
that accord value as perceived in the human world.
Humans, societies, real assets, fiat currencies, liquidity,
legal regimes and regulations
all contribute to form the notion of ``value'' as attached to (bound to)
the constructs (e.g. coins, tokens) that circulate in
the blockchain system,
and which are in-turn implemented
by systems and subsystems at the mechanical layer.
Included also is the notion of legal governance rules
which support humans in making decisions
regarding the operations of a given blockchain as an autonomous system.

\end{itemize}

We believe this general two-level view is consistent with the 
end-to-end principle of the Internet architecture
because it places the human semantics (value) and social interactions
at the ends of (i.e. outside) the mechanical systems.
Interoperability at the mechanical level is necessary for
interoperability at the value level
but does not guarantee it.
Human agreements (i.e. legal contracts) must be used at the value level
to provide semantically compatible meanings
to the  constructs (e.g. coins, tokens) that circulate in
the blockchain system.
Thus, {\em semantic interoperability} is required not only at the value level
but also at the mechanical level.

The mechanical level plays a crucial role in providing technological solutions
that can help humans in quantifying risk through the use
of a more measurable {\em technical trust}~\cite{TPM2003Design}.
In most cases technical-trust is obtained through 
a combination of demonstrably strong cryptographic algorithms,
proper key management,
tamper-resistant hardware (to a specific measure of cost)
and {\em roots of trust} that combine trustworthy computing principles
with {\em legal trust} (e.g. contract that binds the root of trust with
legal enforceable obligations and warranties).

Legal trust is the bridge between the mechanical level
and the value level.
That is, technical-trust and legal-trust support {\em business trust}
(at the value level) by supporting real-world
participants in quantifying and managing risks associated
with transactions
occurring at the mechanical level.
Standardization of technologies that implement technical trust
promotes the standardization of legal contracts 
-- also known as legal trust frameworks -- which in turn reduces the
overall business cost of operating autonomous systems.

\begin{figure}[!t]
\centering
\includegraphics[width=1.0\textwidth, trim={0.0cm 0.0cm 0.0cm 0.0cm}, clip]{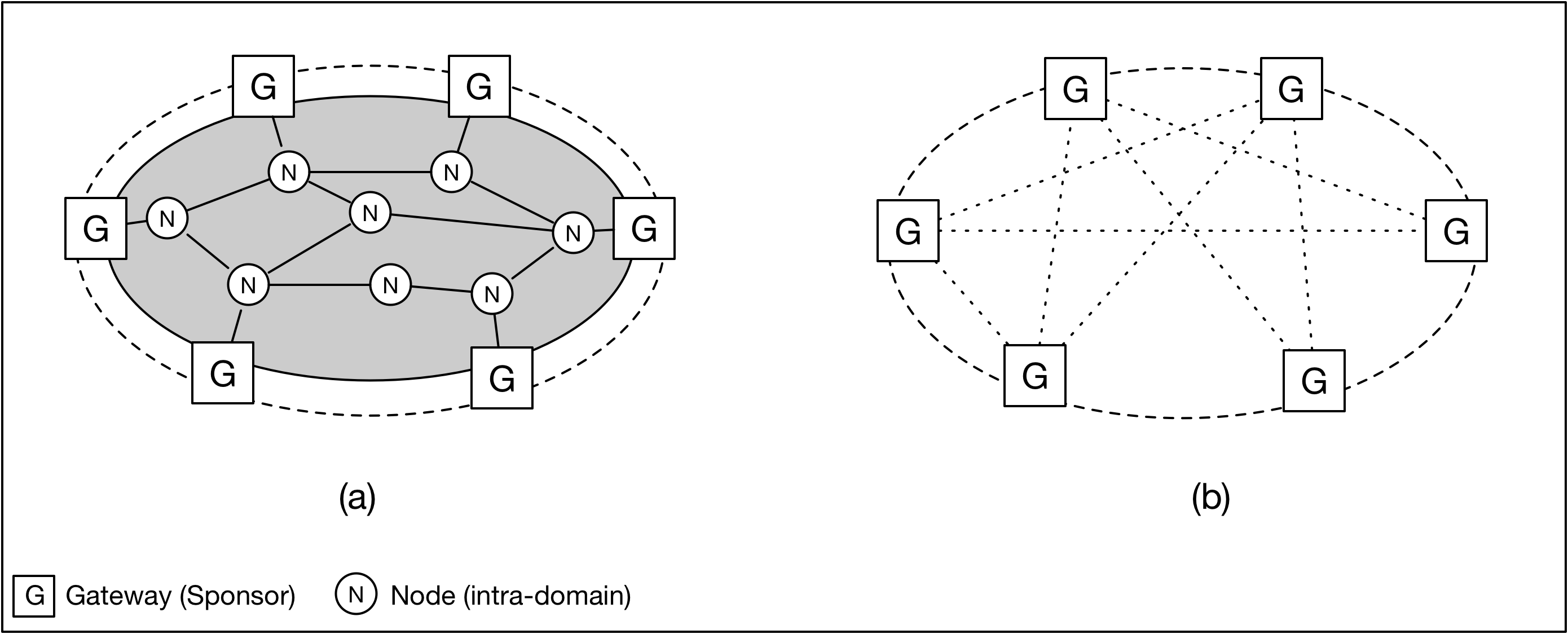}
	%
	%
\caption{Tradecoin model: (a) Blockchain as Autonomous System, and (b) Gateways}
\label{fig:dtcgateway-concept}
\end{figure}

\subsection{The Tradecoin Autonomous System}

The Tradecoin model views
each blockchain system as being a fully fledged autonomous system
in the sense of the Internet architecture.
The basic Tradecoin system is shown in Figure~\ref{fig:dtcgateway-concept}(a).
The blockchain autonomous system consists
of entities which operate intra-domain
(e.g. P2P nodes) and entities that operate inter-domain.

\begin{itemize}
\item	{\em Intra-domain entities}:
The entities dealing with intra-domain transactions
are the {\em peer-to-peer nodes} ($N$).
This includes
nodes that participate in consensus computations (e.g. full mining nodes in Bitcoin~\cite{Bitcoin}),
nodes that ``orchestrate'' consensus computations 
(e.g. Orderers and Endorsers in Fabric~\cite{AndroulakiBarger2018}),
and nodes which perform validations only 
(e.g. Validators in Ripple~\cite{SchwartzYoungs2014}).
In the Tradecoin interoperability model,
an autonomous system may contain multiple blockchain systems.

\item	{\em Inter-domain entities}:
The entities dealing specifically with inter-domain transactions
are denoted as {\em Gateways} and shown as $G$ in 
in Figure~\ref{fig:dtcgateway-concept}(b).
Depending on implementation,
gateways may operate collectively as a group
or they may operate loosely-coupled.

\end{itemize}

In the Tradecoin interoperability model,
the {\em boundary} (perimeter) of the autonomous system is largely determined by
(i) the degree of identity-anonymity and authentication
of the intra-domain entities such as nodes (i.e. to each other in the same domain);
and
(ii) degree of permissionability of the blockchain autonomous system
as seen by foreign entities
(see Section~\ref{subsec:varietybc}).
Thus, unlike classical corporate network topologies
which define a clear (defensive) physical network perimeter,
in blockchain autonomous systems the ``perimeter''
is defined by the participation of the entities
in an intra-domain arrangement.
These intra-domain entities need not be 
located in the same physical proximity,
but in some cases may be required to enter into a legal agreement
(e.g. system rules of operation).
Depending on the use-case,
these entities may be owned by a single organization
(e.g. private corporation),
or be jointly-owned by a multiple organizations (e.g. consortium).

In defining permissionability
the Tradecoin interoperability model uses a number
of permissioning configurations.
The first two pertain to the P2P nodes of the blockchain system,
while the remaining two pertain to the end-users and applications:
\begin{itemize}

\item	{\em Node-permissioned}: In this permissioning configuration,
nodes must be permissioned to participate in one or more
aspects of the operation of the blockchain system.
Using the Hyperledger Fabric~\cite{AndroulakiBarger2018} as an example,
an instance of a Fabric blockchain may require
all the Orderers and Endorsers nodes to be
authenticated and authorized to operate.

\item	{\em Consensus-permissioned}: In this permissioning configuration
only the nodes that participate directly in consensus algorithm
computations need to be authenticated and authorized.

\item	{\em User write-permissioned}: This permissioning configuration
pertains to the end-users and their applications.
In a write-permissioned configuration,
the user/application must be authenticated and authorized
to transmit a ledger-modifying transaction to the blockchain.

\item	{\em User read-permissioned}:
In a read-permissioned configuration,
the user/application must be authenticated and authorized
to read the contents of the ledger of the blockchain system.

\end{itemize}

\begin{figure}[!t]
\centering
\includegraphics[width=1.0\textwidth, trim={0.0cm 0.0cm 0.0cm 0.0cm}, clip]{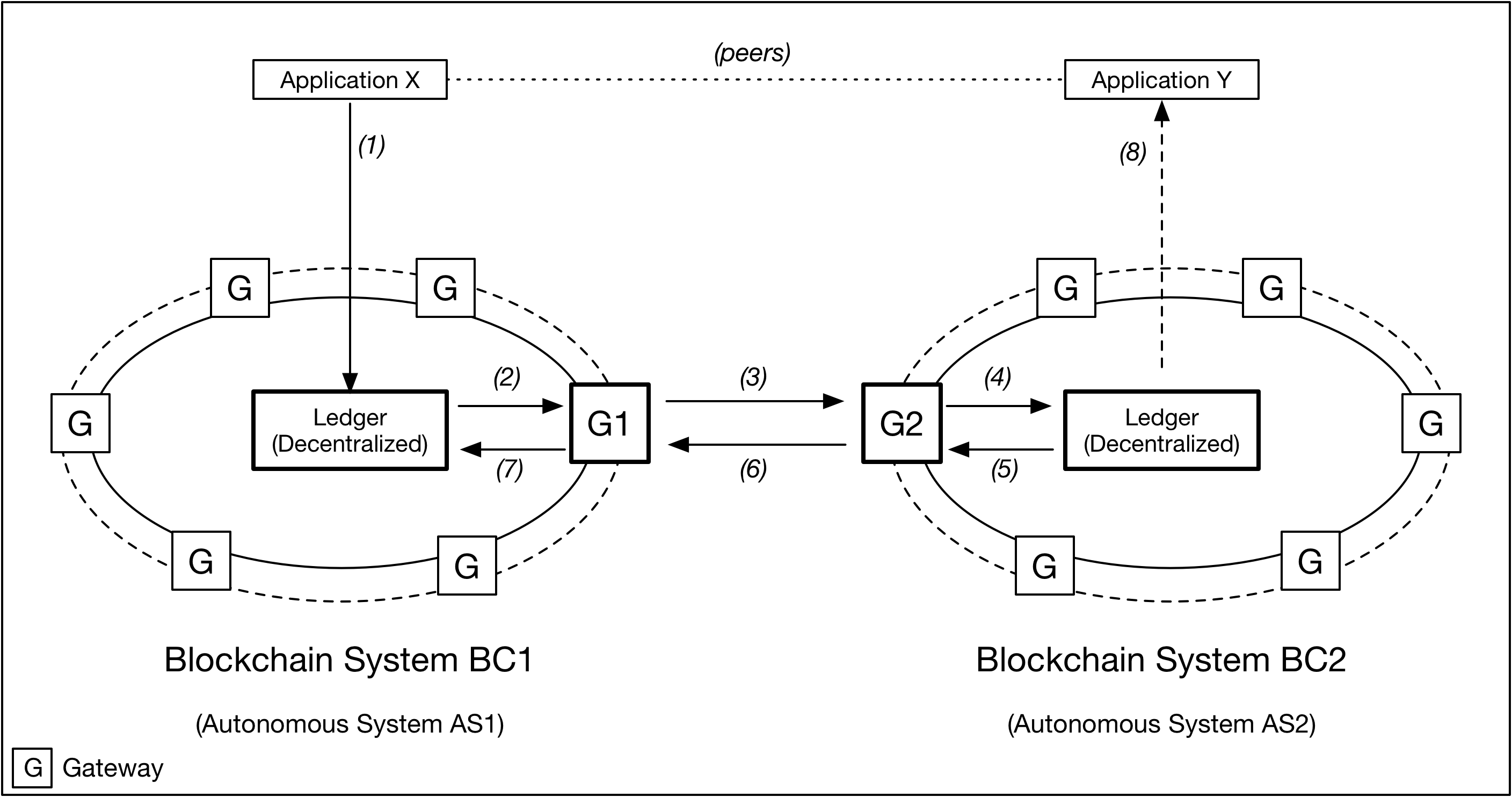}
	%
	%
\caption{Set of Gateways for Reachability and Transaction Mediation}
\label{fig:bc-mediation}
\end{figure}

\subsection{The Role of Gateways}
\label{subsec:RoleGateways}

The set of gateways $G$ collectively provides at least 
three (3) types of functional
support to the entities in the Tradecoin autonomous system:
\begin{itemize}

\item	{\em Value stability}: For implementations of Tradecoin
that exchange value-carrying constructs
(e.g. digital coins, assets),
the gateways collectively act to provide stability at the value level.

Thus, for example, in the case of a Narrow Bank scenario~\cite{TradecoinCapco2018}
which distinguishes asset-contributing entities (called ``sponsors'') from users,
each sponsor may own and operate equal numbers of nodes and
gateways within the shared blockchain system.
The gateways may collectively implement an asset-tracking blockchain,
while the remaining intra-domain nodes implement
the local digital currency of the Narrow Bank.
The value-translation between the coins circulating
in the currency blockchain (intra-domain) against
other foreign denominations
is performed by the gateways, which may hold a basket of real-world assets
(e.g. basket of oil, gold, commodities, greenback, etc)
to back the intra-domain digital currency.

\item	{\em Reachability}: The gateways collectively support reachability
to data intra-domain on ledger of the blockchain.
They support cross-domain lookup
mechanisms from foreign entities (e.g. nodes in other blockchains)
seeking to locate data pertaining to transactions confirmed on the
local ledger.
That is, the gateways provide reachability so that
internal confirmed transaction can be  meaningfully
referenced outside the blockchain system.

\item	{\em Transaction mediation}: 
The gateways provide a mediation function for cross-domain transactions
involving two or more (permissioned) blockchain systems,
where transaction data in ledgers may be considered
as private and sensitive information.

\end{itemize}

\subsection{Reachability}

An interoperable architecture for blockchain systems
must allow not only for entities to be uniquely identifiable and authenticated,
but also for transactions on the ledger
to be uniquely {\em referenceable}
across domains regardless of permissioning configurations:
\begin{itemize}

\item	{\em Endpoint identifier resolutions}: 
Gateways provide a perimeter/boundary
for permissioned blockchain systems for the purposes of naming/identifier resolution.
Thus when a transaction-identifier is externally referenced and resolved
to a ledger-entry inside a permissioned blockchain system,
the gateways ``intercept'' that resolution request
and act as (defensive) {\em end-points} for these external requests.

\item	{\em Identifier masking for data privacy}: Gateways provide ``masking'' (re-naming)
of transaction identifiers in local ledgers inside the blockchain autonomous system.
This may involve mapping different identifiers for the same transaction data,
where one identifier is used for intra-domain referencing 
and another is used for cross-domain referencing.
The gateways also play a role in filtering
information leaving the boundary of a blockchain system,
limiting privacy leakage.

Note that this mapping idea itself is not new
and is deployed today at a global scale in Internet addressing
(e.g. IPv4/IPv6 address mapping, NAT address traversal~\cite{rfc2663}).

\item	{\em Identifier referencing and de-referencing service}:
Gateways may collectively implement a transaction-identifier resolution service,
in similar sense to the hierarchical arrangement
of the Domain Name Service (DNS and DNS-SEC)~\cite{rfc882,rfc2535}.

\end{itemize}

With regards to addressability,
one promising approach
is that of the Inter-Ledger Protocol~\cite{ILP2016} (v1.0.0)
which proposes the use of an addressing scheme
that allows ledgers and nested-ledgers to be identified,
and which proposes a global allocation scheme
for these addresses.

\subsection{Inter-Domain Transaction Mediation}

Gateways in a blockchain autonomous system
may collectively provide a mechanism
to proactively {\em mediate} cross-domain transactions
involving two blockchain systems as distinct autonomous systems.
The gateways in the respective systems must interact
facilitate the atomic and correct recording
of cross-ledger transactions.

Figure~\ref{fig:bc-mediation} illustrates
the simple use-case discussed earlier.
Here, a user with Application X
has his or her asset-ownership (e.g. land title deed)
recorded on the ledger inside blockchain BC1.
The user wishes to transfer legal ownership
of the asset to a different user running Application Y
and have the asset recoded authoritatively on the ledger
inside blockchain BC2.
Both blockchain BC1 and BC2 are permissioned/private
blockchain systems.
Thus, none of the gateways in BC1 can read/write to BC2,
and none of the gateways in BC2 can read/write to BC1.

The term ``authoritative'' here means that
(i) henceforth any external de-referencing of the asset identifier
must resolve to blockchain BC2;
and
(ii) the ledger in BC1 must be ``marked''
for that asset such that it henceforth points to BC2
for any local look-ups for that asset transaction data.

The Tradecoin interoperability model proposes the use
of the classical cross-domain {\em delegation} paradigm
that is well-established in several sectors of
industry (e.g. cross-domain directory services~\cite{rfc1510,MS-KILE}).
Without going into details, 
in Figure~\ref{fig:bc-mediation} the set of gateways
in BC1 and BC2 respectively act as
delegates to the users/applications involved.
Thus,
the entities intra-domain within BC1 and BC2
must trust their gateways to create a temporary {\em bridge}
that allows their gateways to be synchronized temporarily
until both ledgers (in BC1 and BC2 respectively)
have completed recording the asset-transfer transaction.

Since blockchains BC1 and BC2 are permissioned
and one side cannot see the ledger at the other side,
the gateways of each blockchain must ``vouch''
that the transaction have been confirmed on the respective ledgers.
That is, the gateways must issue a legally-binding
signed assertions that makes them liable
for misreporting (intentionally or otherwise).
The signature can be issued by one gateway only,
or it can be a collective {\em group signature}
of all gateways in the blockchain system.
Various cryptographic schemes have been proposed
over the past two decades around group-oriented
signatures~\cite{AtenieseTsudik1999}.
Each of these gateway signatures must also be recorded
on the respective ledgers.

There are several desirable features of the gateway-mediated approach
outline above:
\begin{itemize}

\item	{\em Verifiability of local confirmations}:
Both the transmitter and recipient applications
must be able to independently verify 
that the transaction was confirmed 
on their respective blockchains,
with sufficient data to allow
post-event auditing.

\item	{\em Legally binding signatures of gateways}:
Delegated gateways must have signatures that are binding,
independent of how the gateway(s) was chosen
to be the delegate for the given cross-domain transaction.
Numerous proposals exist for leader-elections,
group voting, consensus and so on.

\item	{\em Equivalent reliability}: 
There must be multiple reliable ``paths'' 
(i.e. set of respective gateways)
between blockchains BC1 and BC2.
Thus, looking at Figure~\ref{fig:bc-mediation}
there must be multiple paths from BC1 and BC2,
and from BC2 to BC1.
Any gateway in BC1 must be able to ``pair'' with any gateway in BC2.

The gateways themselves must never become
a hindrance to completing the a cross-domain transaction.
Attacking gateways must not yield better results (to the attacker)
compared to attacking the P2P nodes intra-domain in the blockchain system.

\item	{\em Global resolution to the correct authoritative blockchain}:
Any external entities seeking to lookup/resolve
an identifier (e.g. linked to the asset)
must always resolve to the correct authoritative blockchain system.
In order words, in Figure~\ref{fig:bc-mediation}
after the cross-domain transaction has completed,
subsequent lookups of the asset must resolve to BC2 (or the gateways of BC2).

\item	{\em Identifiability and authenticity of gateways}:
In order for gateways to act as a delegate for the user/application,
they must be identifiable (i.e. non-anonymous) both
internally (intra-domain) and externally (inter-domain).
Gateways must be able to mutually authenticate each other
without any ambiguity as to their identity, legal ownership or the 
``home'' blockchain autonomous system which they
exclusively represent.

\end{itemize}

\subsection{Peering Agreements for Blockchain Systems}

A key aspect in the Internet architecture that 
promotes and expands the interconnectivity of the autonomous systems
is the {\em peering} agreements between these systems.
In the context of IP routing,
peering is voluntary and occurs typically between Internet Service Providers (ISPs).
The peering agreements
are contracts that define the various interconnection mechanical aspects
(e.g. traffic bandwidth, protocols, etc)
as well as fees (``settlements'') and possible penalties.
Peering is made possible by the standardization of 
cross-domain routing protocols
(e.g. BGP~\cite{rfc1105})

Historically, a peering arrangement can be considered
 ``public'' in the sense that there is a ``group contract''
among a group of ISPs that allow any group-member
to transit traffic through another member.
Peering agreements can also be ``private'' in
that it is entered into by two ISPs,
providing mutually better service levels to both parties.
Peering agreements provides the ISPs with the correct incentive structure
for them to operate their autonomous system as a business.

For the interoperability of blockchain systems,
a notion similar to peering and peering-agreements must be developed
that: 
(i) define the semantic compatibility required for two blockchains
to exchange cross-domain transactions;
(ii) specifies the cross-domain protocols required;
(iii) specifies the delegation and technical-trust mechanisms to be used;
and
(iv) define the legal agreements (e.g. service levels, fees, penalties, liabilities, warranties) for peering.

It is important to note that in the Tradecoin interoperability model,
the gateways of a blockchain system
represents the peering-points of the blockchain.
The peering agreement can be established either
with another blockchain system (private bilateral),
with a group of blockchain systems (private multi-lateral),
or with an open ``exchange'' system (public peering).

\section{Discussion: Survivable Digital Currency and CBDC}
\label{subsec:survivable-CBDC}

Several governments in the world have recently indicated
their interest in using digital currency technology
as the basis for a future reserve currency,
also referred to as {\em Central Bank Digital Currency} (CBDC)
or {\em Sovereign Money}~\cite{ShengGeng2018}.
Some examples include Switzerland~\cite{Vollgeld2017}, 
China~\cite{Knight2017} and Russia~\cite{GalouchkoBaraulina2017}.

Although digital currencies may enhance economic strengths of certain nations,
the move to a digital currency comes with its own set of challenges.
As nations increasingly rely on digital infrastructures,
those infrastructures -- if not designed and operated correctly --
may bring a different set of liabilities.

In this section we pose a number of questions
concerning the survivability of digital currencies and CBDCs,
particularly in the face future sophisticated cyber-attacks
on the monetary electronic infrastructure.
Instead of focusing on the survival of one blockchain system,
we discuss {\em communities} of blockchain autonomous systems
as the basic unit of concern,
recognizing that there will be many communities
with varying sizes, technical capabilities and varying 
instruments of value.
\begin{itemize}

\item	{\em Independence of communities of blockchain systems}:
Assuming that blockchain systems deployments
evolve organically in a similar manner to the Internet,
what would be the most useful composition of ``communities''
of blockchain autonomous systems from the point of view of currency survivability?
A key aspect is the ability of a community to continue
functioning economically while the monetary flows
into (out of) that community are temporarily disrupted.

Today the Internet operates not only at the level of local ISPs,
but also across long-haul physical networks coast to coast,
and overseas.
Connectivity of the Internet within the US and Canada will most likely continue
to operate even if overseas connectivity was lost.
Similarly, loss of coast to coast IP connectivity
will still allow local ISPs to offer services to its 
local physically connected communities.
Multiple reliable IP traffic paths coast to coast
ensures that disruptions from attacks
have minimal effects.

\item	{\em Survivable peering models}:
What kinds of peering agreements need to be developed
for blockchain autonomous systems within a community to enhance
the operational survival chances of that community. 
Furthermore, should {\em super-peering} agreements be developed
for cross-community engagements that come into effect
in emergency situations.

\item	{\em Self-protecting blockchain autonomous systems}:
How can advanced artificial intelligence (AI) and machine learning (ML)
technologies be used to enhance the protection
of communities of blockchain autonomous systems?
Distributed AI/ML tools can be used to analyze 
community behaviors on the transactional level
and provide insight into anomalies that may indicate
unauthorized attempts to alter or influence 
monetary flows within a given community.
The predictive capabilities of these tools could
be used to support the enhanced shaping of monetary-flows in anticipation
of emergent attacks.

\item	{\em Cross-community recovery from attacks}:
How can communities of blockchain autonomous systems
re-establish transactional connectivity automatically and organically
(at the mechanical-level and value-level)
after they have experienced ``isolation'' due to successful attacks?
Furthermore, how can ``old infrastructure'' (e.g. transactional systems,
interbank networks, paper cash, etc) be used to boot-up
communities into a stable state (preferably into the same
pre-attack state).

\end{itemize}

\section{Conclusions}

The fundamental goals underlying the Internet architecture
has played a key role in determining the interoperability
of the various networks and service types,
which together compose the Internet as we know it today.
A number of design principles emerged from the evolution
of internet routing in the 1970s and 1980s,
which ensured the scalable operation of the Internet
over the last three decades.

We believe that a similar design philosophy is needed
for interoperable blockchain systems.

The recognition that a blockchain system is
an autonomous system is an important
starting point that allows notions
such as reachability, referencing of transaction data in ledgers,
scalability and other aspects to be understood more meaningfully  --
beyond the current notion of throughput (``scale''),
which is often the sole measure of performance
stated with regards to some blockchain system.

Furthermore, interoperability
forces a deeper re-thinking into how permissioned and permissionless
blockchain systems can interoperate
without a third party (such as an exchange).
A key aspect is the semantic interoperability
at the value level and at the mechanical level.
Interoperability at the mechanical
level is necessary for interoperability 
at the value level but does not guarantee it.
The mechanical level plays a crucial role in 
providing technological solutions that can help
humans in quantifying risk through the use of 
a more measurable notion of technical-trust.
Human agreements (i.e. legal contracts) 
must be used at the value level to 
provide semantically
compatible meanings to the constructs (e.g. coins, tokens) 
that circulate in the blockchain system.



\end{document}